\documentclass[fleqn,usenatbib]{mnras}
\usepackage{newtxtext,newtxmath}

\usepackage[T1]{fontenc}
\usepackage{ae,aecompl}

\usepackage{graphicx}	
\usepackage{amsmath}	
\usepackage{footnote}
\usepackage{tablefootnote}
\usepackage[hyphenbreaks]{breakurl}

\newcommand{\src}{4U 1724--30}

\title[Probing NS LMXB \src]{Probing spectral and
temporal evolution of the neutron star low-mass X-ray binary 4U 1724--30 with {\em AstroSat}
}

\author[U. Kashyap et al.]{
Unnati Kashyap$^{1}$\thanks{E-mail: phd1801121005@iiti.ac.in},
Manoneeta Chakraborty$^{1}$,
and Sudip Bhattacharyya$^{2}$
\\
$^{1}$DAASE, Indian Institute of Technology Indore, Khandwa Road, Simrol, Indore - 452020, India\\
$^{2}$Department of Astronomy and Astrophysics, Tata Institute of Fundamental Research, 1 Homi Bhabha Road, Colaba, Mumbai - 400005, India\\
}

\date{Accepted XXX. Received YYY; in original form ZZZ}

\pubyear{2020}

\begin{document}
\label{firstpage}
\pagerange{\pageref{firstpage}--\pageref{lastpage}}
\maketitle

\begin{abstract}

We report the broadband spectro-temporal study of the poorly studied accreting neutron star (NS) low mass X-ray binary (LMXB) 4U 1724--30 using data from Soft X-ray Telescope (SXT) and Large Area X-ray Proportional Counters (LAXPC) instruments on board {\em AstroSat}. The dim persistent LMXB source was observed with {\em AstroSat} over 4 epochs in 2017, all of which corresponded to a low-luminosity non-thermal emission dominated (hard/island) emission state with modest spectral evolution. All the X-ray broadband spectra can be modelled by a combination of thermal emission from the NS boundary layer (BL) or NS surface and a non-thermal emission component possibly originating from the inverse Comptonization of the disc seed photons. We investigate the presence of frequency and energy-dependent variabilities to probe the origin of the disc/coronal fluctuations. We also report the detection of a Type-I X-ray burst displaying a  photospheric radius expansion (PRE). During the burst, a hard X-ray shortage in the 30-80 keV energy band and the enhancement of the persistent emission reveal the burst feedback on the overall accretion process. Using the touch-down burst flux $\sim$ $4.25 \times 10^{-8}$  erg s$^{-1}$ cm$^{-2}$, the distance of the source is estimated as $\sim$ 8.4 kpc.

\end{abstract}



\section{Introduction}



Low mass X-ray binaries (LMXBs) are composed of a neutron star (NS) or a black hole that accretes matter from a low-mass companion star via Roche-lobe overflow. These systems are ideal laboratories for studying the physics of accretion and signatures in the strong-field regime \citep{2006PhDT.........6D, 2000LNP...556..173K}. Investigations of NS LMXBs can specifically provide vital clues on the nature of ultra-dense degenerate matter \citep{2010AdSpR..45..949B, 2016ARA&A..54..401O, 2018Degenaar....457}.

  A significant population of NS LMXBs is composed of persistently accreting X-ray sources, known as persistent LMXBs, while sources that become active intermittently are labeled as transient LMXBs. LMXB transients show a wide variation in X-ray luminosity ranging between $10^{-8}-1$ times the Eddington luminosity ($L_{Edd}$). A certain level of variability in the accretion rate has been observed even for the persistent LMXBs. With varying accretion rates, NS LMXBs show distinct states, exhibiting characteristic evolution of the source spectral and timing properties \citep{BARRET2001307,2004astro.ph.10551V,2007A&ARv..15....1D,2007ApJ...667.1073L}. 
As the X-ray luminosity increases, the source can progress from a low-luminosity non-thermal emission-dominated hard state to an intermediate state to finally a thermal emission-dominated soft state. At relatively lower accretion rates, the disc could be truncated relatively far away from the central compact object \citep{1999MNRAS.303L..11Z, 2001AAS...198.1001E, 2004AIPC..714...52K, 2004ApJ...601..439T}. In such a hard state, the emission can be modelled by a single temperature blackbody representing the thermal emission from the boundary layer or the thin disc and a hard component arising from the inverse-Comptonization by a hot plasma in the inner accretion flow or the corona with the latter dominating \citep{2000ApJ...544L.119D, 2016ApJ...827..134S}. In the soft state, however, the disc contribution becomes generally stronger. 
 
  Along with the long-term spectral state evolution, the NS LMXBs also exhibit short and rather unpredictable episodes of intense variation in the X-ray luminosity in the form of Type-I X-ray bursts. Type-I (thermonuclear) bursts in NS LMXBs occur due to the unstable burning of hydrogen/helium at the surface of the accreting neutron star in low-mass X-ray binaries (LMXBs). During the thermonuclear ignition, if the luminosity of the source reaches its Eddington limit, the NS envelope is proposed to be pushed outwards due to radiation pressure leading to the expansion of the NS photosphere. Consequently, the size of the emitting areas increases with a relative drop in blackbody temperature. Once the photosphere cools following the expansion, it settles back down, receding to its original radius at a stage called the touch-down. This corresponds to the decrease in the emitting area with the increase in the inferred temperature. After the touch-down phase, the characteristic cooling is observed. This phenomenon of temperature drop and rise with the expansion and the contraction of the radius near the peak of the burst is defined as Photospheric Radius Expansion (PRE) \citep{2018SSRv..214...15D, 2002A&A...382..947K,1980ApJ...240L.121G,1983PASJ...35...17E,1984ApJ...277L..57L,1984ApJ...276L..41T}.

4U 1724--30 is a dim accreting NS LMXB which is located in the core of the globular cluster Terzan 2 \citep{1998A&A...339..802G}. The distance of this cluster is 7.5--12 kpc \citep{1996yCat.7195....0H}. The low luminosity ($\sim 10^{37}$ erg s$^{-1}$) indicates the dim nature of 4U 1724--30  \citep{article}.
 Following the classification of NS LMXBs by \cite{1989A&A...225...79H}, the X-ray colour-colour diagram of this source was observed to follow a trend consistent with that of atoll sources. In its high flux hard state, a peculiar, long-lasting ($\sim$ 300 d) flaring event was reported from this NS LMXB \citep{2018MNRAS.477.3353T}. The source was observed to be variable in X-rays on a timescale of 10 years in the 1990s, and the reason behind such variability was reported to be either due to the evolution of the donor star or due to the influence of a third star \citep{2002AstL...28...12E}. Variabilities at different frequencies including  kiloHertz (kHz) Quasi-Periodic Oscillation (QPO) have also been reported  for this NS atoll source \citep{2008ApJ...687..488A,1998A&A...333..942O, 1999NuPhS..69..245B}. The detections of PRE bursts from 4U 1724--30 have also been reported \citep{2008ApJS..179..360G,2000A&A...357L..41M,2011ApJ...742..122S} previously.

The main focus of this paper is to study the broadband emission behavior of the poorly studied NS LMXB 4U 1724--30. We also probe the evolution of different thermal and non-thermal components in the accretion disc-boundary layer/NS surface and the corona over the four observational epochs starting from 2017 February 27 to 2017 July 29. The broadband capability of {\em AstroSat} enables spectro-temporal study in both soft and hard X-ray regimes simultaneously ( see Section \ref{section4.1} \& \ref{section4.3}). Additionally, we report the time-resolved spectroscopy of the Type-I (thermonuclear) X-ray burst detected during the last observation and the burst influence on the enhancement of the persistent emission and the hard X-ray shortage. As the detected burst shows Photospheric Radius Expansion (PRE), we use the touch-down flux to estimate the distance to the binary system. The enhanced timing capability of Large Area X-ray Proportional Counters (LAXPC) on board {\em AstroSat} provides scope for time-resolved spectroscopic study of thermonuclear bursts, which may provide an insight into the burst-accretion interaction processes ( see Section \ref{section4.2}).

\section{Observation and Data Reduction}
\label{section2}

\begin{table*}
\centering
\caption{Observation details (LAXPC) of 4U 1724--30 including observation number, orbit number, dates of observations, start and stop times of observations as well as effective on source exposure in seconds for all the four observational epochs (see Section \ref{section2.2}).
\label{tab1}}

\begin{tabular}{|c|c|c|c|c|c|c|c|c|c|c|c|c|}
\hline
Observation number & Orbit number & Date (dd-mm-yyyy) & Start time (hh:mm:ss) & Stop time (hh:mm:ss) & Exposure (ks)  \\ \hline
 1 &7671-7680 & 27-02-2017 &07:20:44.52 &23:54:11.16&59.607\\
2 & 7902-7903 & 14-03-2017 - 15-03-2017  & 22:29:59.22 &02:14:44.78&13.486\\
3 & 9723-9731 & 15-07-2017 - 16-07-20  & 23:36:21.36&13:54:19.59&51.478\\
4 & 9924-9933 & 29-07-2017 - 30-07-2017& 16:29:59.77& 05:39:39.25&47.803\\
\hline
\end{tabular}
\end{table*}

\begin{table*}
\centering
\caption{Observation details (SXT) of 4U 1724--30 including observation number, orbit number, dates of observations, start and stop times of observations as well as effective on source exposure in seconds for all the four observational epochs (see Section \ref{section2.1}).
\label{tab2}}

\begin{tabular}{|c|c|c|c|c|c|c|c|c|c|c|c|c|}
\hline
Observation number &Orbit number & Date (dd:mm:yyyy) & Start time (hh:mm:ss) & Stop time (hh:mm:ss) & Exposure (ks)  \\ \hline
 1 & 7671-7680&27-02-2017  & 07:25:57.45  &  22:19:25.33 & 10.504 \\ 
2 & 7902-7911&15-03-2017 &   00:15:58.98   &  13:29:50.69 & 8.3562\\
3 & 9724-9731&16-07-2017 &  00:46:59.37  & 13:46:00.98  &  12.175 \\
4 & 9924-9932& 29-07-2017 & 16:29:42.91 & 04:00:28.85  & 11.180\\
\hline
\end{tabular}
\end{table*}


The LMXB source 4U 1724--30 was observed simultaneously with SXT and LAXPC on four epochs: 2017 February 27 (ObsID: 9000001058), 2017 March 15 (ObsID: 9000001082),  2017 July 15 (ObsID: 9000001386), and 2017 July 29 (ObsID: 9000001416) respectively. The observation details for LAXPC and SXT are listed in Table~\ref{tab1} and Table~\ref{tab2}, respectively.

 \subsection{LAXPC}
 \label{section2.2}
 
The Large Area X-ray Proportional Counters \footnote{\url{https://www.tifr.res.in/~astrosat_laxpc/astrosat_laxpc.html}} (LAXPC) instrument consists of three identical proportional counter detector units named LAXPC10, LAXPC20, and LAXPC30 with the largest effective area in the mid-X-ray range 3--80 keV. It is one of the primary payloads on {\em AstroSat} with a large area collection of $\sim 6000$ cm$^{2}$. The deadtime of these detectors is $\sim$ 42 $\mu$s and they have a time resolution of 10 $\mu$s which makes LAXPC ideal for fast timing analysis \citep{2017CSci..113..591Y}.

 \texttt{LaxpcSoft} software is used to analyze LAXPC data, process the light curves, spectra, and extract background and response files, respectively. The light curves and spectra are obtained using Level 2 event files \citep{2017ApJS..231...10A}. LAXPC30 data are not useful due to gas leakage caused by the gain instability, and we do not use LAXPC10 data due to low gain. All LAXPC observations are taken in the Event Analysis (EA) mode. In the Event mode, the arrival times and energies of each photon are recorded. Barycenter correction is applied to the LAXPC level 2 data using the \texttt{as1bary} tool.

 \subsection{SXT}
\label{section2.1}

The Soft X-ray Telescope\footnote{\url{https://www.tifr.res.in/~astrosat_sxt/index.html}} \citep[SXT; ][]{10.1117/12.2235309,2021JApA...42...17B} on board {\em AstroSat} is capable of providing spectra in the energy range 0.3--8 keV, soft X-ray images, spectroscopy as well as variability observations of cosmic sources by focusing X-rays on a cooled Charge Coupled Device (CCD) \citep{2017JApA...38...29S}. The effective area of SXT is 90  cm$^{2}$  at 1.5 keV. It has a focal length of 2 meters. 
The time resolution of SXT is $\sim$ 2.4 s in PC mode and $\sim$ 278 ms in FW mode. The telescope field of view, on-axis FWHM, and half-power diameter (HPD) are $\approx$ $40'$, $2'$, and $10'$, respectively. 
 The SXT data are processed using \texttt{sxtpipeline} to generate the filtered level-2 cleaned event files. Then we apply \texttt{SXTEVTMERGERTOOL} to merge the different orbits and produce a single level-2 Event file for each observation epoch. All the SXT event data are in FW (Fast Window) mode, which provides a better time resolution. To extract the images, the light curves, and the spectra, the ftool \texttt{XSELECT}, which is available as a part of the \texttt{Heasoft 6.24} package, is used. The data do not suffer from any pile-up effects given the low source intensity, the FW mode, and the wide PSF of SXT\footnote{\url{https://www.tifr.res.in/~astrosat\_sxt/instrument.html}} \citep{10.1117/12.2235309}. A circular region of 5 arcmin radius is considered as a source region around the source location at RA of $261.884^{\circ}$ and DEC of $-30.812^{\circ}$ \citep{2006ApJ...636..765B}. During spectral analysis, the response file
(sxt\_pc\_mat\_g0to12.rmf), the ancillary response file  (sxt\_fw\_excl00\_v04\_20190608.arf) and, the blank sky background spectrum file
(SkyBkg\_comb\_EL3p5\_Cl\_Rd16p0\_v01.pha) provided by the SXT team are used.   

\section{Results}
\label{section3}
\subsection{Spectral Analysis}
\label{section3.1}

\begin{figure}
    \centering
    \includegraphics[width=0.45\textwidth]{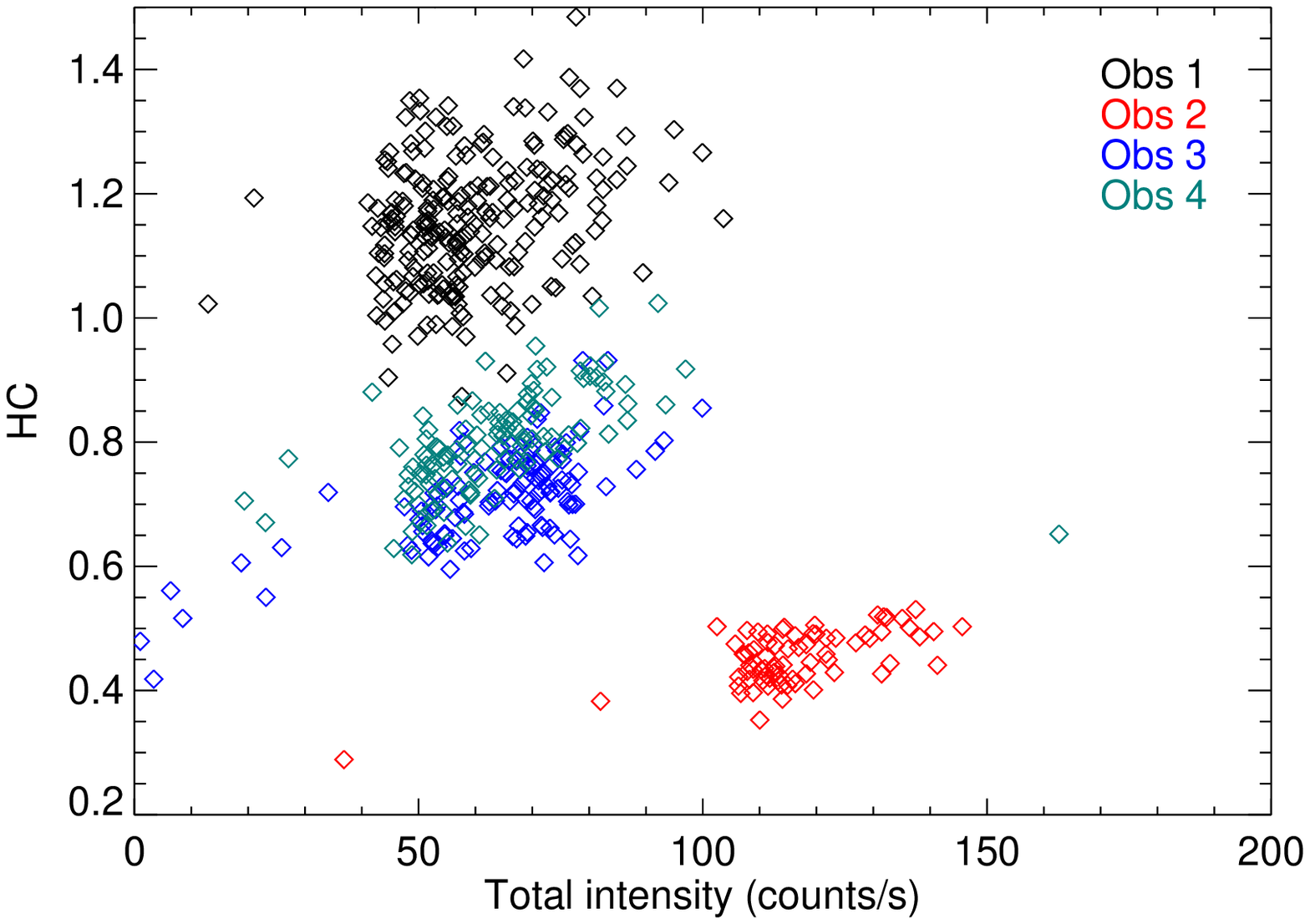}
    \caption{Hardness-Intensity Diagram of 4U 1724--30 for observation 1 (February 27), observation 2 (March 14), observation 3 (July 15) and observation 4 (July 29) (LAXPC). Hard colour is defined as the 16.0-9.7 keV/9.7-6.0 keV background subtracted count rate ratio and intensity as the whole energy band background subtracted count rate. In each individual point, 100 s exposure is averaged (see Section \ref{section3.1}). } 
     \label{hid}
\end{figure}
\begin{figure*}
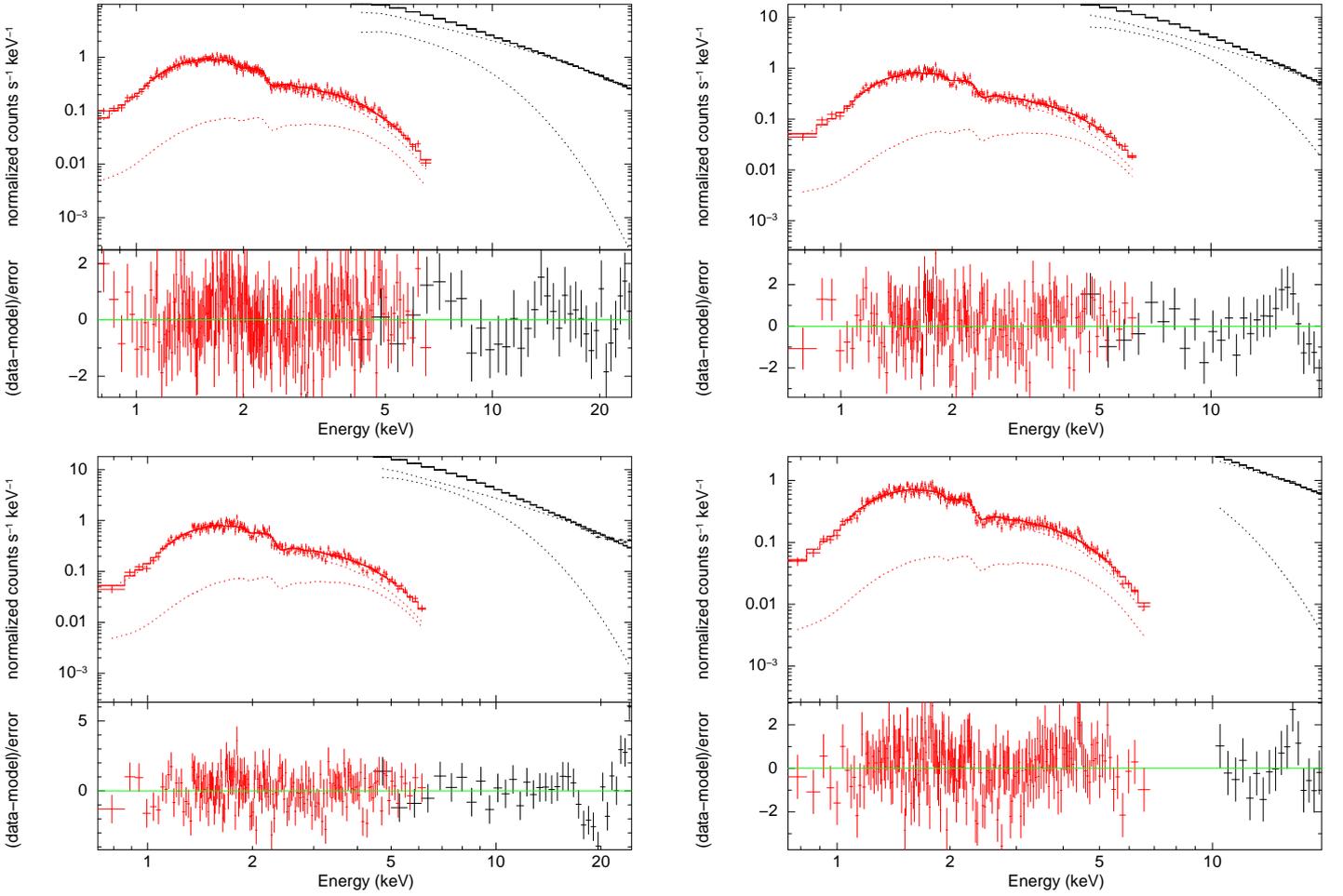

    \hspace*{-1.4cm}
    \begin{tabular}{lr}
    \includegraphics[width=0.35\textwidth,angle =270]{Figure2a.ps} &
    \includegraphics[width=0.35\textwidth,angle =270]{Figure2b.ps} \\
    \includegraphics[width=0.35\textwidth,angle =270]{Figure2c.ps} &
    \includegraphics[width=0.35\textwidth,angle =270]{Figure2d.ps} \\
    \end{tabular}
    \caption{Simultaneous broadband Energy spectrum of 4U 1724--30 (left to right, top to bottom) as observed from SXT (red) and LAXPC (black) on board {\em AstroSat} for the observation 1 (February 27), observation 2 (March 14), observation 3 (July 15) and observation 4 (July 29) and the best fitting model consisting of absorbed blackbody radiation and comptt component {\tt const*tbabs*(bbodyrad+comptt)}. The individual additive components (blackbody radiation and comptt) are also shown by the dotted curves (see Section \ref{section3.1}). } 
    \label{spec_obs3}
\end{figure*}

\begin{figure}
   \centering
   \includegraphics[width=0.48\textwidth]{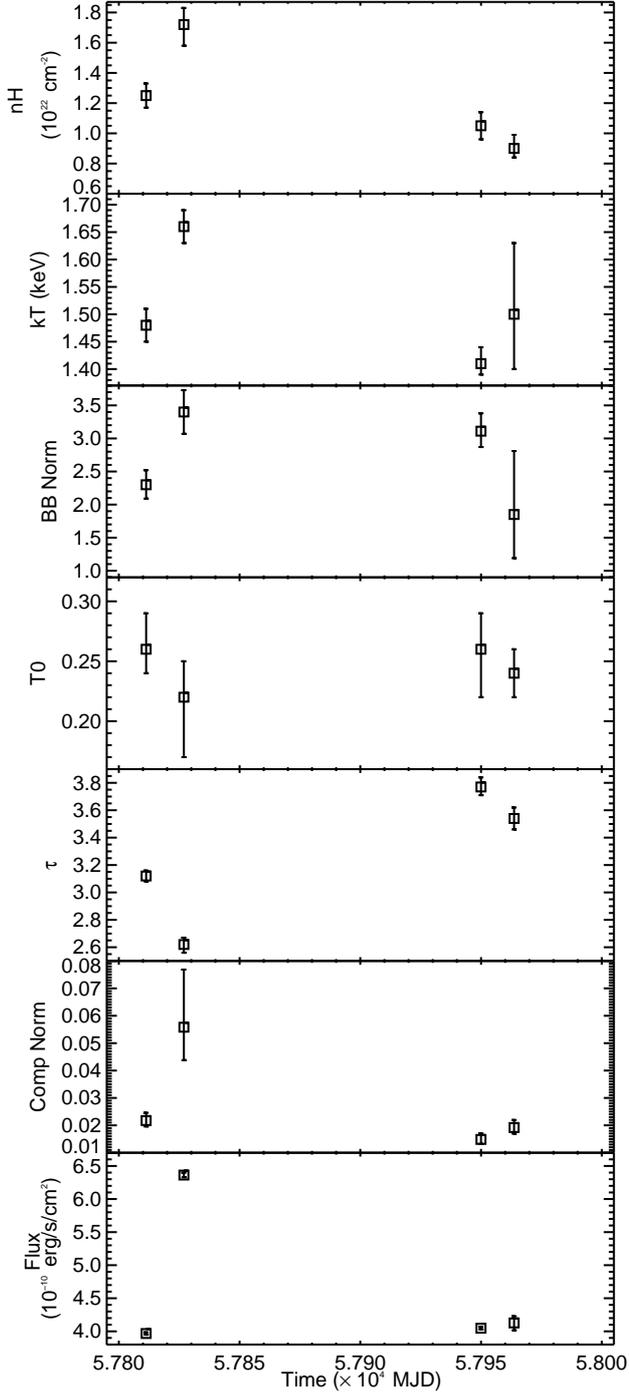}   
   \caption{Evolution of the best fit spectral parameters in all four observational epochs of 4U 1724--30. First panel: galactic neutral hydrogen column density ($n_H$), second panel: blackbody temperature $kT$, third panel: blackbody normalization, fourth panel: input soft photon temperature, fifth panel: plasma optical depth, sixth panel: Comptonization normalization, seventh panel: source flux in the 4-25 keV energy band for the observation 1 (February 27), observation 2 (March 14), observation 3 (July 15) and observation 4 (July 29) corresponding to the best fitting model. The source flux, $kT$, $n_H$, blackbody and Comptonization normalization reach maximum near observation 2 where the source was in a relatively softer, higher flux state. 1 $\sigma$ errors are quoted on the parameters (see Section \ref{section3.1}).}
   \label{para_evol}
\end{figure}

\begin{table*}
\centering

\caption{Best fitting spectral parameters corresponding to the best fitting model {\tt const*tbabs*(bbodyrad+comptt)} for the observation 1 (February 27), observation 2 (March 14), observation 3 (July 15), and observation 4 (July 29) of 4U 1724--30. \label{tab:specpar} (see Section \ref{section3.1})} 
\label{spec_para}
\begin{tabular}{|c|c|c|c|c|c|c|c|c|c|c|}
\hline
 Obs No. & $n_H^{a}$ & $kT^{b}$ & BB Norm$^{c}$ &$T0^{d}$&$kT_{e}^{e}$&$\tau^{f}$&Comp Norm$^{g}$  & LAXPC Flux $^{h}$ &  $\chi^2$ (DOF)  \\
& $10^{22}$ cm$^{-2}$ &keV  & &keV&keV &&$10^{-2}$&$10^{-10}$  $\mathrm{erg\,s^{-1} cm^{-2}}$ &  \\
\hline
1 &$1.25^{+0.08}_{-0.08}$ &$1.48^{-0.03}_{+0.03}$ & $2.30^{+0.21}_{-0.21}$&$0.26^{+0.02}_{-0.02}$  &8.0 (fixed)  &$3.12^{+0.04}_{-0.04}$&$2.17^{+0.28}_{-0.22}$&$3.95^{+0.02}_{-0.02}$&0.95 (255)  \\
\\
2&$1.72^{+0.10}_{-0.13}$  &$1.66^{+0.03}_{-0.03}$ &$3.40^{+0.33}_{-0.33}$ &$0.22^{+0.04}_{-0.04}$  &8.0 (fixed)  &$2.62^{+0.06}_{-0.06}$&$5.58^{+0.02}_{-0.01}$ &$6.36^{+0.04}_{-0.03}$&1.32 (199) \\
\\
3 &$1.05^{+0.08}_{-0.09}$  & $1.41^{+0.03}_{-0.03}$&$3.11^{+0.27}_{-0.24}$ &$0.26^{+0.03}_{-0.03}$  &8.0 (fixed)  &$3.77^{+0.06}_{-0.06}$&$1.49^{+0.22}_{-0.16}$&$4.05^{+0.02}_{-0.02}$&0.98 (215)  \\
\\
4& $0.90^{+0.10}_{-0.05}$ &$1.50^{+0.13}_{-0.10}$ &$1.85^{+0.96}_{-0.66}$ & $0.24^{+0.02}_{-0.02}$ & 8.0 (fixed) &$3.54^{+0.08}_{-0.08}$&$1.92_{-<0.01}^{+<0.01}$&$4.13^{+0.10}_{-0.11}$&1.22 (222)  \\

\hline

\end{tabular}
\begin{flushleft}
\footnotesize{$^a$ Neutral hydrogen column density}\\
\footnotesize{$^b$ Blackbody Temperature } \\
\footnotesize{$^c$ Blackbody Normalizations}\\
\footnotesize{$^d$ Input soft photon (Wien) temperature }\\
\footnotesize{$^e$ electron temperature (keV)}\\
\footnotesize{$^f$ Plasma optical depth }\\
\footnotesize{$^g$ Comptonization normalization} \\
\footnotesize{$^h$ Flux in the 4.0-25.0 keV energy band (LAXPC) }\\
\end{flushleft}
\end{table*}

We examine the joint spectral fitting of LAXPC20 and SXT to study the broadband emission spectra of the source during all four observations. The spectral fitting and statistical analysis are done using the \texttt{XSPEC v 12.10.0c} spectral fitting package distributed as a part of the \texttt{Heasoft 6.24} package. As mentioned in the SXT user manual, a systematic error of 2\% must be included in the spectral fits, and systematic error for LAXPC spectral data was reported to be 2\% \citep{Leahy_2019, 2017ApJS..231...10A, 10.1093/mnras/stz1327}. Therefore, for the joint fitting, a 2\% systematic error is added to the spectrum to account for the instrumental uncertainty in the response. We ignore photon counts below 0.7 keV and above 7 keV for SXT and below 4 keV and above 25 keV for LAXPC, to avoid background effects and loss of sensitivity. Due to low counts, SXT spectra are grouped using {\tt GRPPHA} to have a minimum count rate of 50 counts per bin.
We perform the joint spectral fit of SXT and LAXPC20 data with different single and multi-component models. The emitted spectrum is modified by the presence of neutral hydrogen absorption in the interstellar medium, and this is taken care of by using {\tt Tbabs} model. The abundances and photoelectric cross-sections are adopted from \cite{2000ApJ...542..914W}. For the purpose of joint spectral fitting, we add a {\tt constant} component to all the models to represent the cross-calibration constant between the LAXPC20 and the SXT instruments. No single-component model gives an acceptable fit in all cases ($\chi^2 \gtrsim 2 $). We have tried two-component models such as blackbody and power-law, blackbody and thermal Comptonization, blackbody and Comptonization, disc blackbody and thermal Comptonization, disc blackbody and Comptonization model, disc blackbody and empirical power-law model to fit the continuum spectra.  Among these, we find the blackbody plus Comptonization model ({\tt const*tbabs*(bbodyrad+comptt)} in \texttt{XSPEC}) to be the best fitting model where parameters are physically acceptable with relatively better reduced $\chi^2$ values for all four observations. The  model {\tt bbodyrad} consists of parameters $kT$ and normalization, $N=(R/d_{10})^{2}$ , where R is the blackbody radius in kilometers and $d_{10}$ is the source distance in units of 10 kpc. Whereas, the model {\tt comptt} consists of parameters such as input soft photon temperature, plasma optical depth, and Comptonization normalization. It is to be noted here that the parameter $(kT_{e})$ is completely unconstrained in observations 2 and 4. So, we first fit the joint spectra of observations 1 and 3, where the parameters are found to be well constrained. The value of the electron temperature $(kT_{e})$ obtained from the fitting of these observations was unvarying and found to be $\sim$ 8 keV. We then fix the electron temperature at 8 keV for the joint spectral fitting of all the observations to obtain better constraints on all the parameters. Figure~\ref{spec_obs3} show the joint LAXPC20 and SXT spectra for all the observations fitted with the best-fitting model ({\tt const*tbabs*(bbodyrad+comptt)}) and their corresponding residuals. The origin of the residuals, especially in the LAXPC data, is mainly due to the uncertainties in the instrumental response and background effects. It is to be noted here that for the spectral fitting, we consider the LAXPC spectra in the 4-20 keV and 10-20 keV energy range only for observations 2 and 4, respectively, as in those cases, the instrument systematics is significantly present.
Using the \texttt{XSPEC} model {\tt cflux}, we compute the source fluxes. The best-fitting spectral parameters corresponding to the chosen model ({\tt const*tbabs*(bbodyrad+comptt)}) obtained from the fits are reported in Table~\ref{spec_para}. The temporal evolution of the parameters such as galactic neutral hydrogen column density ($n_H$), blackbody temperature $kT$,  blackbody normalization, input soft photon temperature, plasma optical depth, Comptonization normalization, and the source flux in the 4-25 keV energy band during the four observations is shown in Figure~\ref{para_evol}. The column densities $n_{H}$ for all the four observations are observed to be almost similar to the previously reported value $\sim 1 \times10^{22}$ cm$^{-2}$ \citep{1998tx19.confE.324B}. The galactic neutral hydrogen column density ($n_H$), blackbody temperature $kT$, and Comptonization normalization peaks for observation 2, where the source was in a higher flux (and relatively softer) state as described by the HID (Figure~\ref{hid}). \\

\subsection{Analysis of the thermonuclear burst}
\label{section3.2}

\begin{figure}
    \centering
    \includegraphics[width=0.50\textwidth]{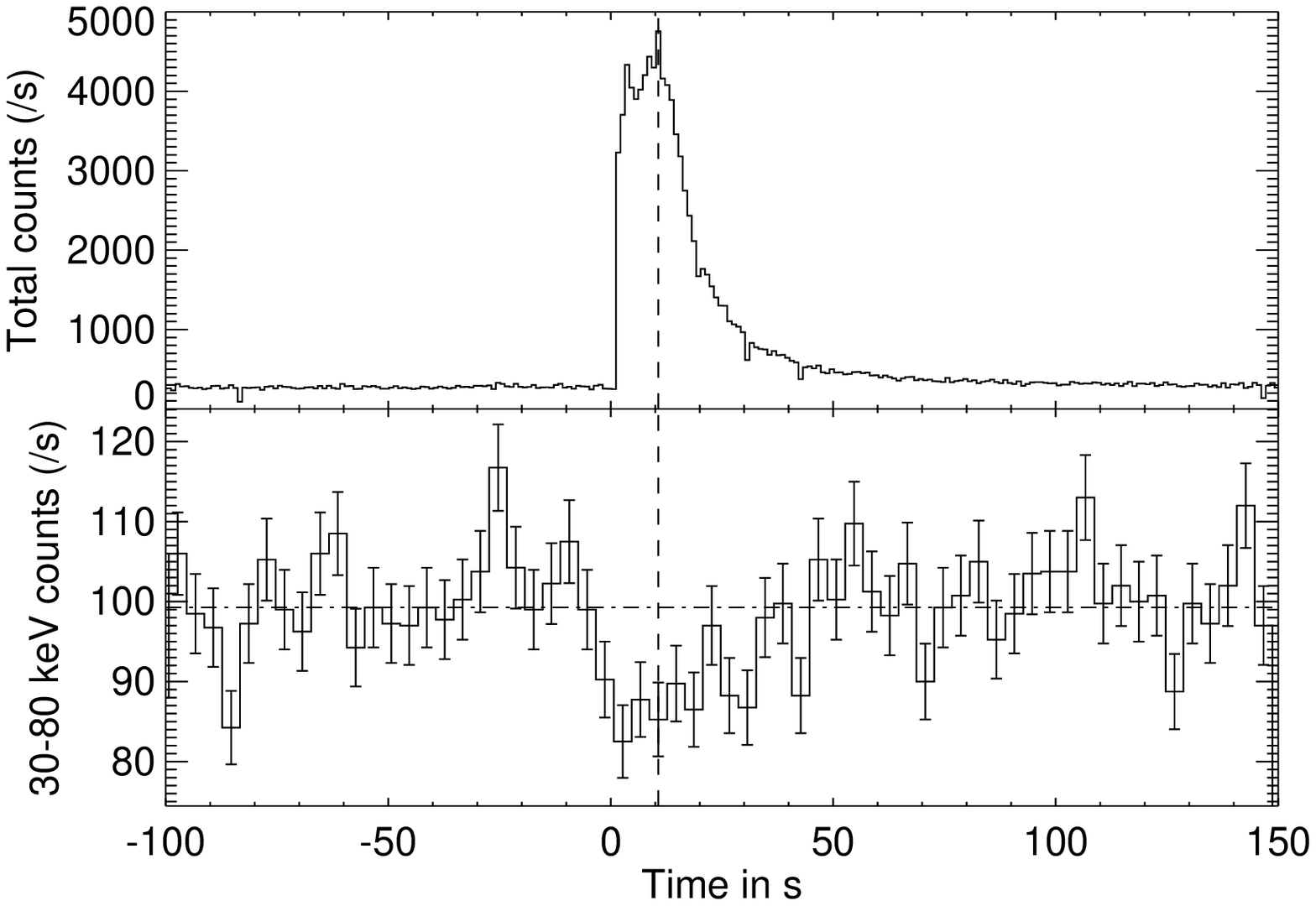}
    \caption{{\it Top panel:} Energy integrated burst light curve (LAXPC) with 1 s binning, {\it Bottom panel:} The simultaneous 30-80 keV light curve with 4 s time binning for better representation. Both light curves span duration starting from 100 s before the detected burst to 150 s after during observation 4 (July 29)  from the accreting NS LMXB 4U 1724--30 (see Section \ref{section3.2}).} 
    \label{burstlc}
\end{figure}

\begin{figure*}
    \centering
    \includegraphics[width=0.45\textwidth]{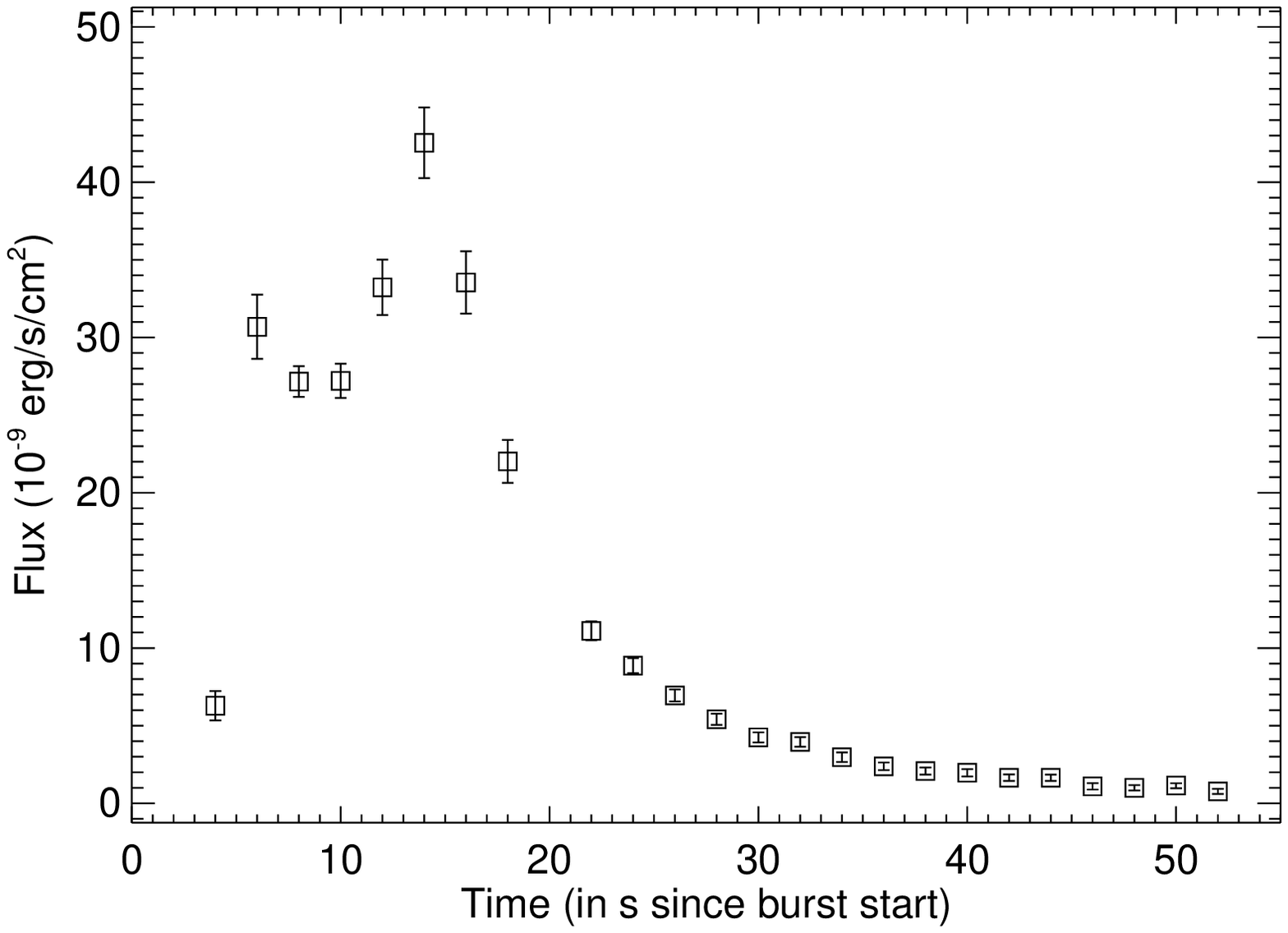}
    \includegraphics[width=0.45\textwidth]{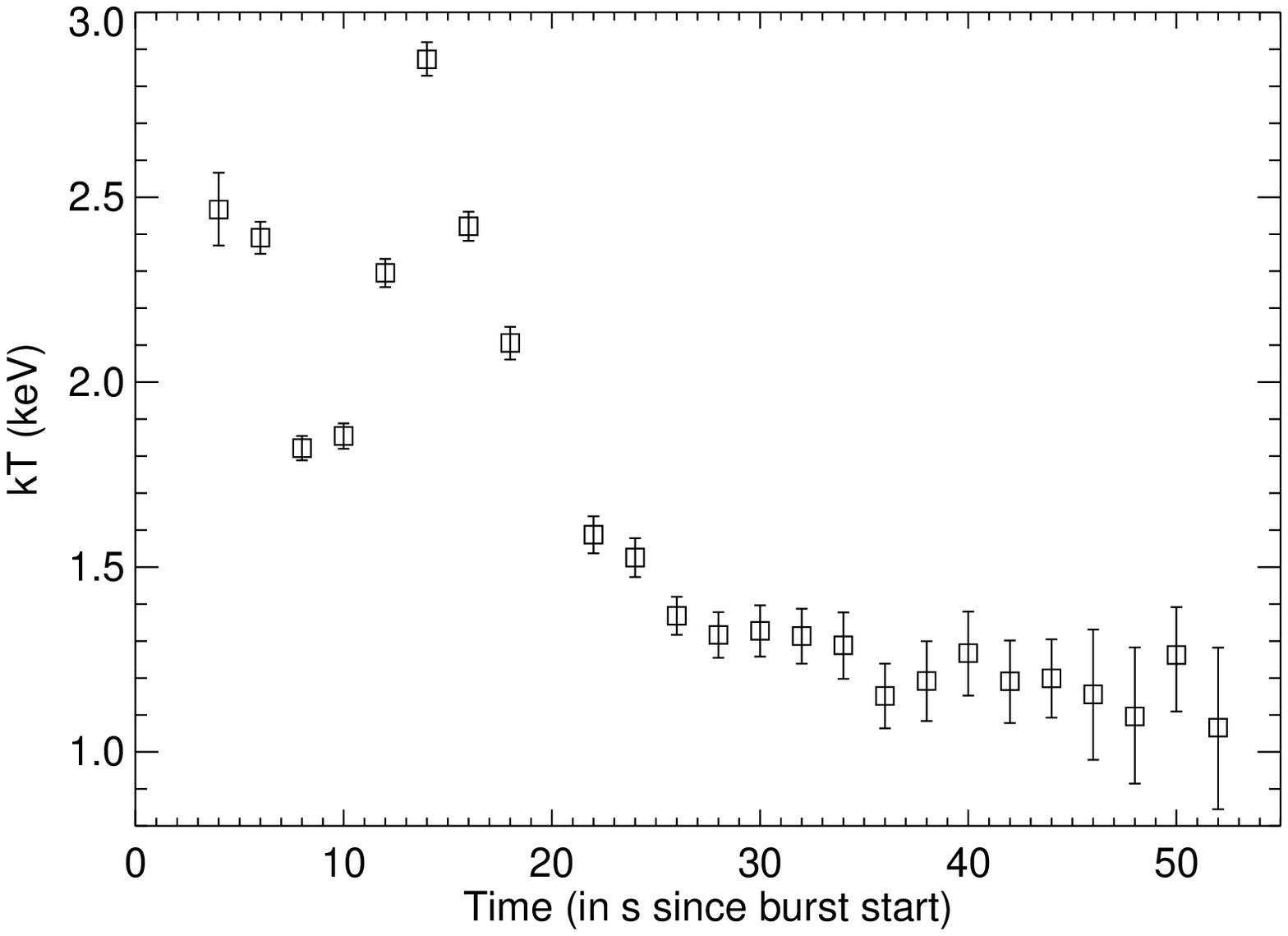}
    \includegraphics[width=0.45\textwidth]{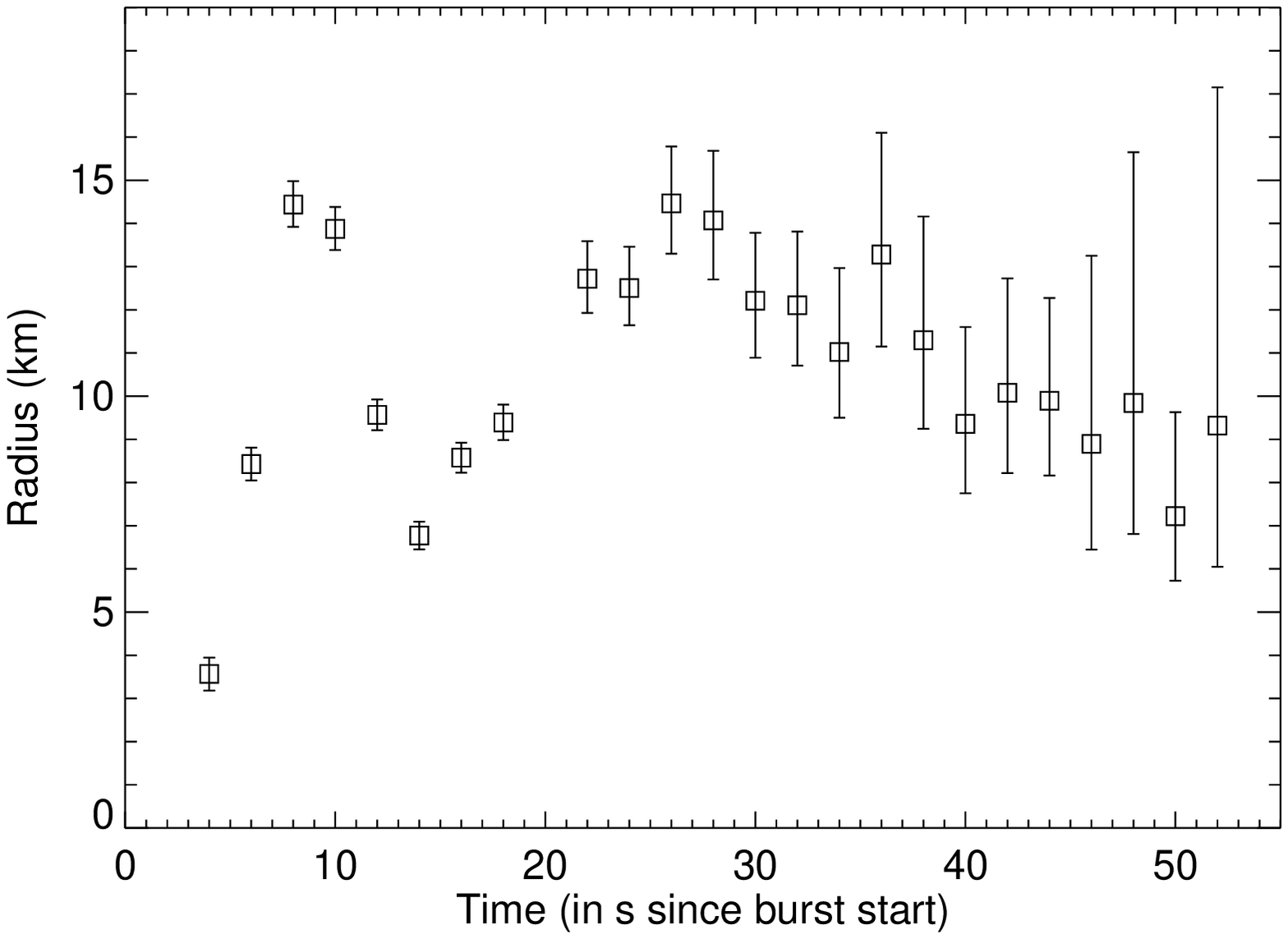}
    \includegraphics[width=0.45\textwidth]{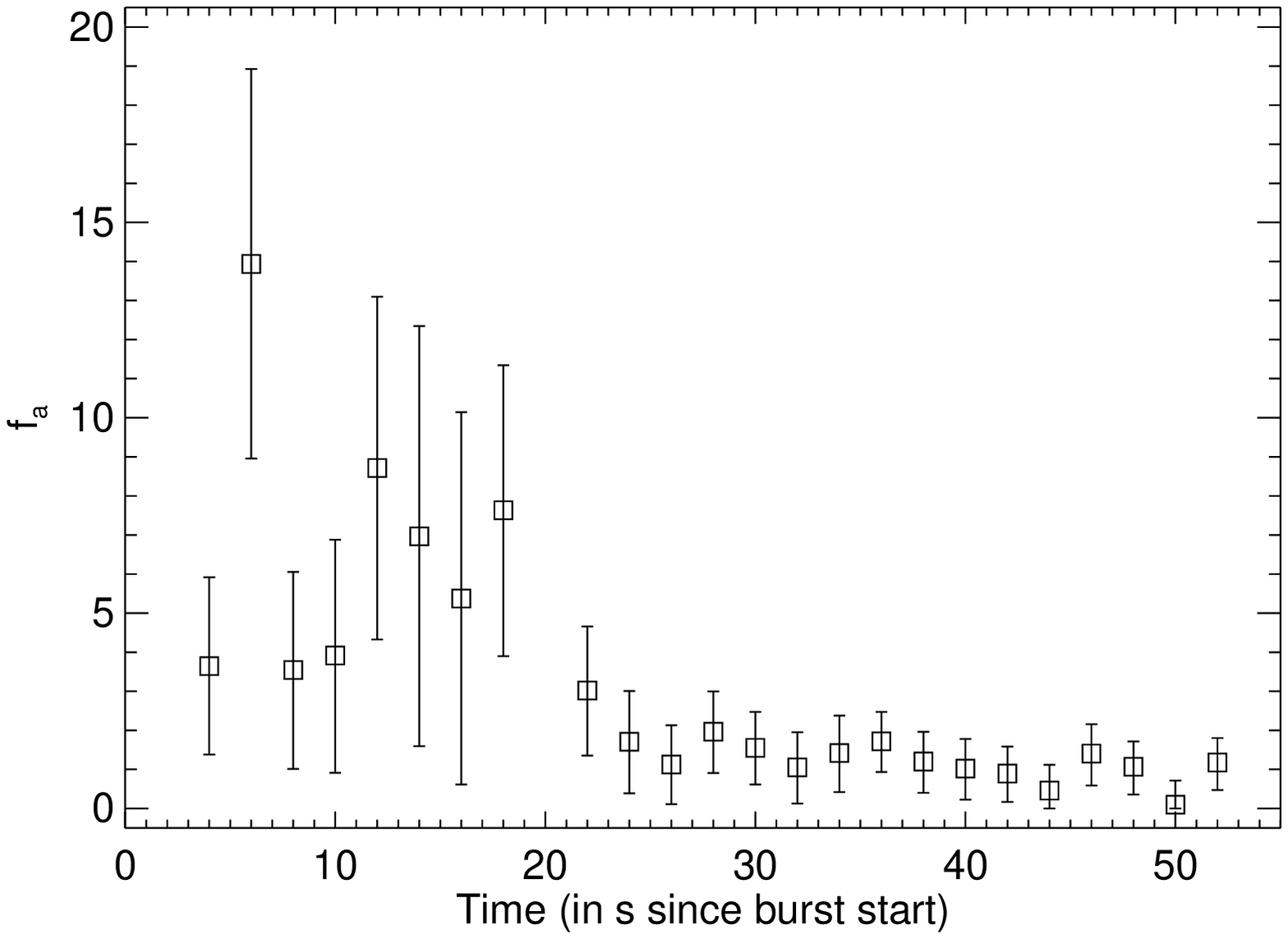}
    \caption{Variation of best fit spectral parameters during the thermonuclear burst detected during observation 4 (July 29) of \src. Evolution of the source flux in the 4-25 keV energy range with 2 s binning (top left) and blackbody temperature $kT$  (top right), blackbody radius (bottom left) and the constant factor $f_{a}$ (bottom right) (see Section \ref{section3.2}).}
    \label{burst_para_evol}
\end{figure*}

We have done a detailed time-resolved spectroscopic analysis of the thermonuclear burst detected in the LAXPC light curve of observation 4. This significant increase in X-ray flux could be observed in the LAXPC data only as it was outside the good time intervals and exposure of SXT. The top panel of Figure~\ref{burstlc} shows the energy integrated burst light curve with 1 s time resolution. The duration of the burst is $\sim 50$ s, and the burst profile is observed to be consistent with a typical thermonuclear burst with double-peaked morphology \citep{2008ApJS..179..360G}. The dashed line demonstrates the time of maximum intensity at the peak of the thermonuclear burst. The bottom panel of Figure~\ref{burstlc} shows the burst light curve in the hard X-ray energy range ($\sim$  30-80 keV). The depression of the hard X-ray flux near the peak of the burst signifies a shortage of high-energy X-ray photons at that time. 

To probe the spectral evolution during the thermonuclear burst, we have extracted spectra of 2 s time intervals from the complete 50 s burst light curve. Due to low counts in the time-resolved burst spectra, the extracted 2 s spectra are grouped using \texttt{GRPPHA} to have a minimum count rate of 10 counts per spectral energy bin.  We fit each 2 s burst spectra using both the conventional ({\tt const*tbabs*bbodyrad} model) method considering the pre-burst emission as background and the $f_{a}$ method to take care of burst-accretion interaction effects \citep{2013ApJ...772...94W}. The burst spectral fits are performed within the 4.0-25.0 keV energy range. In the $f_{a}$ method, we fit the burst spectra taking instrumental background (obtained from the pipeline) as the background with an absorbed blackbody model adding a variable persistent emission component \citep{2013ApJ...772...94W}. Thus, the burst spectra for each 2 s time bin are fitted in \texttt{XSPEC} using a model {\tt const*tbabs*bbodyrad + $f_{a}$*(const*tbabs*(bbodyrad + comptt))}. The component for the burst is  ({\tt const*tbabs*bbodyrad}). The neutral hydrogen column density $n_{H}$ of the burst component is fixed at the continuum value $\sim$ $0.73 \times 10^{22}$ cm$^{-2}$ as obtained from the continuum spectral fitting of observation 4.  
The second term in the model in the $f_a$ method represents the variable persistent emission, where the factor $f_a$ tracks the variation of the persistent emission during the burst, and the parameters of the {\tt const*tbabs*(bbodyrad+comptt)} model are kept fixed at the values obtained from the continuum spectral fitting of observation 4. From the spectral analysis using the  $f_{a}$ method,  the persistent emission is observed to vary during the burst quantified by the evolution of the factor $f_{a}$. 
Thus, we report the results obtained from the time-resolved spectroscopy using the $f_{a}$ method only. 
Figure~\ref{burst_para_evol} shows the evolution of the best fitting parameters such as flux, $kT$, blackbody radius (obtained from blackbody normalization considering a distance of $\sim$ 8.42 kpc which is estimated using the peak PRE burst flux, see Section \ref{section4.2}), and $f_{a}$ obtained from the time-resolved spectroscopic analysis. The errors shown reflect a confidence of 1 sigma. From Figure~\ref{burst_para_evol}, we see the temperature decay with the increase in blackbody radius during the first 6 s of the burst (expansion). Following the expansion phase (after 6 s), the blackbody temperature increases with the drop in blackbody radius (emitting area). The touch-down at which the flux reaches its maximum value ($\sim 4.25 \times 10^{-8}$ $\mathrm{erg\,s^{-1} cm^{-2}}$) is observed at 12 s since the burst start. After 12 s, the burst decays and the thermal cooling is observed (Figure~\ref{burst_para_evol}). Thus, this is an example of a strong PRE burst \citep{2008ApJS..179..360G,2018SSRv..214...15D, Bhattacharyya2021} with distinct expansion and contraction phases. Moreover, the bottom right panel of Figure~\ref{burst_para_evol} shows the presence of a variable persistent emission contribution during the burst and its evolution. It is observed that the value of the multiplicative factor $f_{a}$, representing the fractional increase to the persistent emission contribution, exceeds 1 and remains greater than 1 until $\sim$ 30 s after the burst onset. This may signify the relative increase in the mass accretion rate onto the NS during the burst.

Furthermore, we search for burst oscillations in the entire LAXPC energy range, but no oscillations are detected. The 1$\sigma$ upper limit of RMS amplitude is found to be 2.15\%. 

\subsection{Timing Analysis}
\label{section3.3}

\begin{table}
\centering
\caption{1 $\sigma$ upper limits on any possible periodicity in a range of frequencies 100-500 Hz the observation 1 (February 27), observation 2 (March 14), observation 3 (July 15), and observation 4 (July 29) of 4U 1724--30 (see Section \ref{section3.3}).}
\label{period}

\begin{tabular}{|c|c|}
\hline
Obs No. & 1 $\sigma$ upper limit(\%)\\ 
\hline
1 & 7.66\\

2 & 9.99\\

3 &10.41 \\

4 &10.28 \\

\hline

\end{tabular}

\end{table}

\begin{figure}
    \centering
   \includegraphics[width=0.40\textwidth]{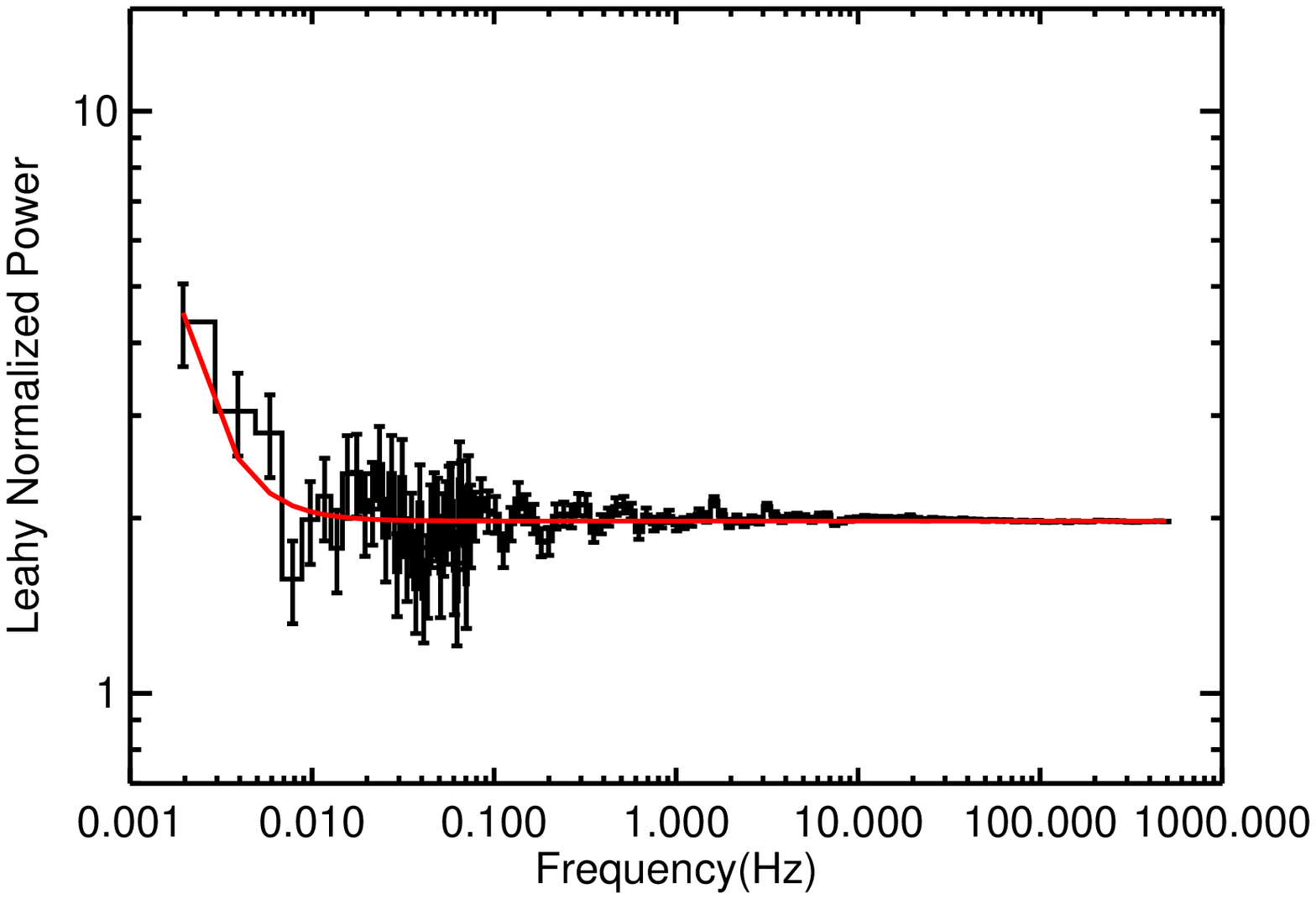}
   \includegraphics[width=0.40\textwidth]{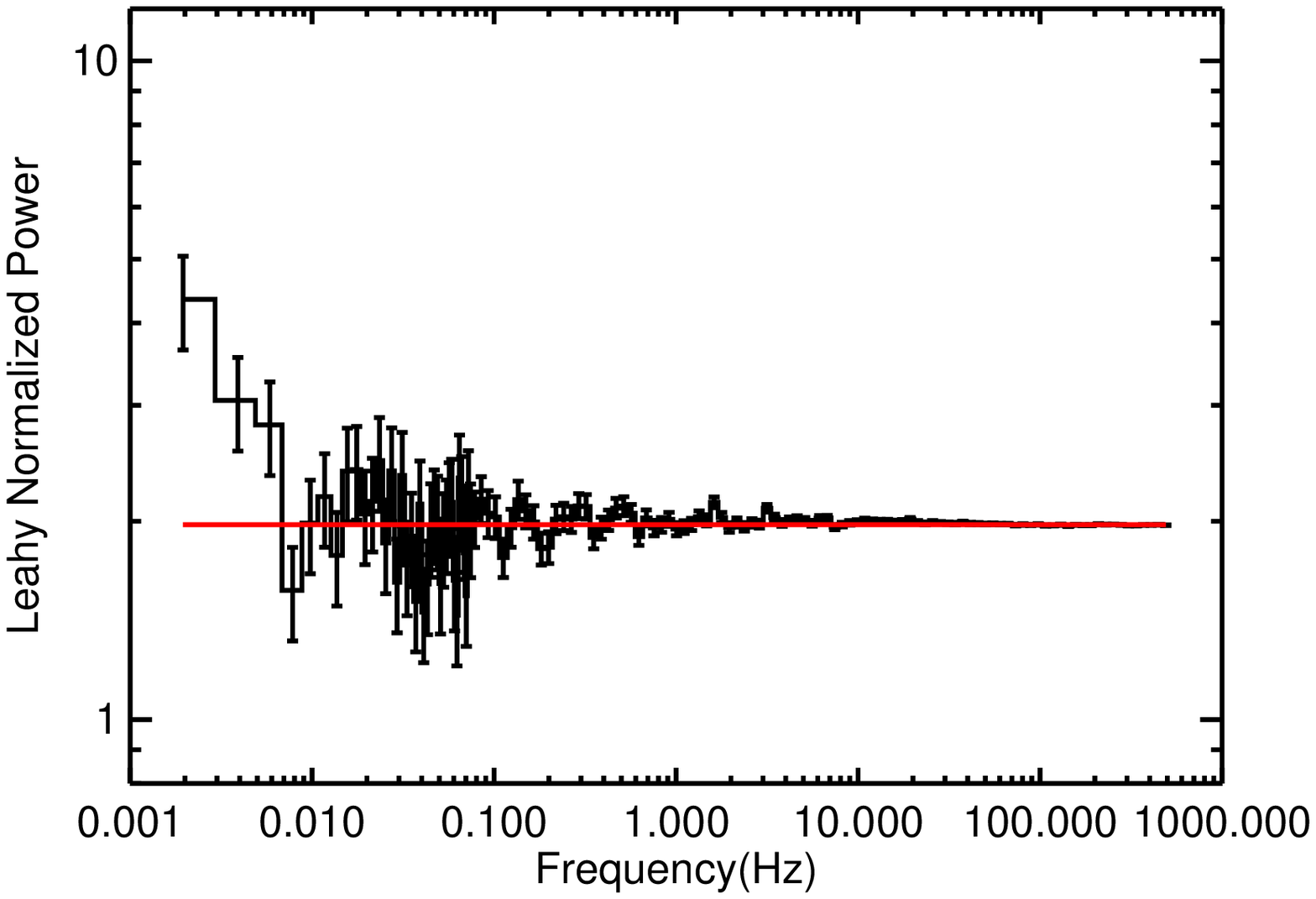}
    \caption{Logarithmically-rebinned (LAXPC20) power spectrum for observation 1  (February 27)  of 4U 1724--30 in the Leahy normalized power representation and the best fitting models: constant+powerlaw (above) and constant model (below) (see Section \ref{section3.3}).} 
     \label{pwspec_lxp}
\end{figure}

\begin{figure}
    \centering
   \includegraphics[width=0.50\textwidth]{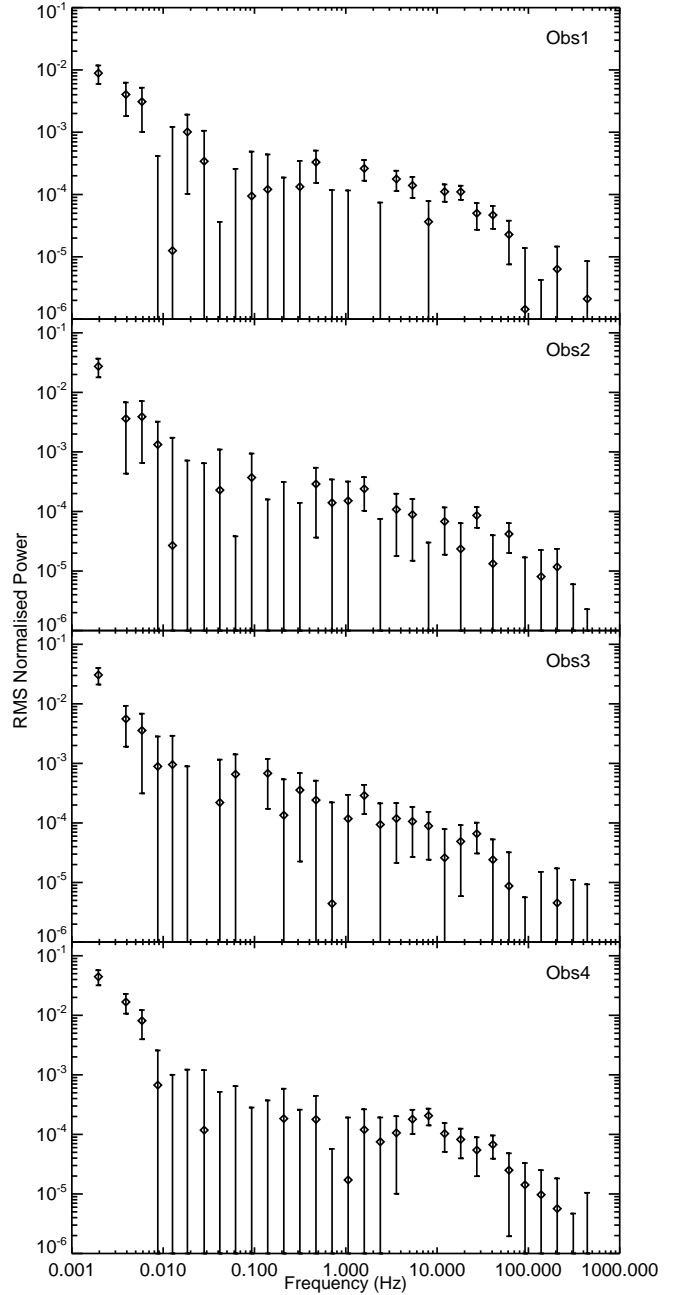}
    \caption{Poisson noise subtracted Logarithmically-rebinned power spectra for all the four (LAXPC20) observations of 4U 1724--30 in the RMS (root mean square) normalized power representation (see Section \ref{section3.3}).} 
     \label{pwspec_lxp_rms}
\end{figure}

\begin{table*}
  \centering
  
  \caption{Best fitting parameters: Leahy normalized power spectra fit to models $(\alpha+\beta \times (f/0.01)^{-\nu})$ and $\alpha$, Degrees of freedom and reduced $\chi^2 $ value (LAXPC 2) for the observation 1 (February 27), observation 2 (March 14), observation 3 (July 15) and observation 4 (July 29) of 4U 1724--30 (see Section \ref{section3.3}). }
  \label{modl_para}
  \renewcommand{\arraystretch}{1.2}
  \begin{tabular}{|p{2cm}|p{3cm}|l|l||p{2cm}|}
    \hline
    \hline
    {\textbf{Observation}} & \textbf{Model} & \multicolumn{2}{|c|}{\textbf{Parameters}} &\textbf{ $\chi^2$ (DOF)} \\
    \cline{3-4}
    &&\hspace{1cm}\textbf{$\alpha$} & \hspace{1cm}\textbf{$\nu$} &  \\
    \hline
   {1}&$(\alpha+\beta \times (f/0.01)^{-\nu})$ &1.98 $\pm<$0.01 & 2.18$\pm$0.80  & 1.78 (128)\\
      &$\alpha$&1.98 $\pm<$0.01&&1.87 (130)\\
   \hline
   {2}&$(\alpha+\beta \times (f/0.01)^{-\nu})$ &1.96 $\pm<$0.01 & 2.68$\pm$0.82  & 1.41 (128)\\ 
   &$\alpha$&1.96 $\pm<$0.01&&1.48 (130)\\
   \hline
   {3}&$(\alpha+\beta \times (f/0.01)^{-\nu})$ &1.98 $\pm<$0.01 & 2.64$\pm$0.73  & 1.21 (128)\\ 
   &$\alpha$&1.98 $\pm<$0.01&&1.31 (130) \\
   \hline
   {4}&$(\alpha+\beta \times (f/0.01)^{-\nu})$ &1.98 $\pm<$0.01 & 2.50$\pm$0.51 & 1.34 (128)\\ 
   &$\alpha$&1.98 $\pm<$0.01&&1.48 (130)\\
   \hline
   \hline
  \end{tabular}
\end{table*}

To investigate the time variability of the NS LMXB source 4U 1724--30 during all the four observations, we have obtained the Leahy normalized power spectra \citep{1983ApJ...266..160L} using data segments (integration time) of 512 seconds from each observation with a sampling time of 1/1024 s such that the lowest available frequency is 1/512 Hz and the Nyquist frequency is 512 Hz. To reduce noise in the Leahy normalized powers, we average the power spectra obtained from all the 512 s segments from the entire observation to obtain resultant power spectra for each observation \citep{1989ASIC..262...27V}. The 1 $\sigma$ upper limits on any possible periodicity obtained in the 100-500 Hz frequency range of all four observations are mentioned in Table~\ref{period}.  To better represent the various variability and noise components present, the power spectra are then binned in logarithmically spaced bins, and the morphology of the logarithmically rebinned power spectra are then analyzed for the presence of any additional features at lower frequencies. We then subtract a Poisson noise estimated from the averaged power between the frequency ranges 150-512 Hz to obtain the Poisson noise subtracted averaged power spectra. For the purpose of timing analysis and plotting, \texttt{IDL version 7.0} is used.

All the generated LAXPC power-spectra are observed to comprise a combination of white noise (uncorrelated constant noise at higher frequencies) and red noise (variable noise at the lower frequencies with a $\sim 1/f^2$ dependence). To study and quantify the timing properties, we consider constant plus power-law model: $\alpha+\beta \times (f/0.01)^{-\nu}$ as well as constant model: $\alpha$. Here $\alpha$, $\beta$, and $\nu$ are the parameters of the model, and $f$ is the frequency of variability. All the power spectra are observed to be satisfactorily modelled by the constant plus power-law model ($1/f^\nu$ noise). The best-fitting parameters for both the models for all the four observational epochs are given in Table~\ref{modl_para}, and the logarithmically rebinned power spectra for observation 1 (LAXPC20) with the fitted models are shown in Figure~\ref{pwspec_lxp}. Figure~\ref{pwspec_lxp_rms} shows the Poisson noise subtracted logarithmically rebinned power spectra for all four observations.

\begin{figure}
    \centering
    \includegraphics[width=0.45\textwidth]{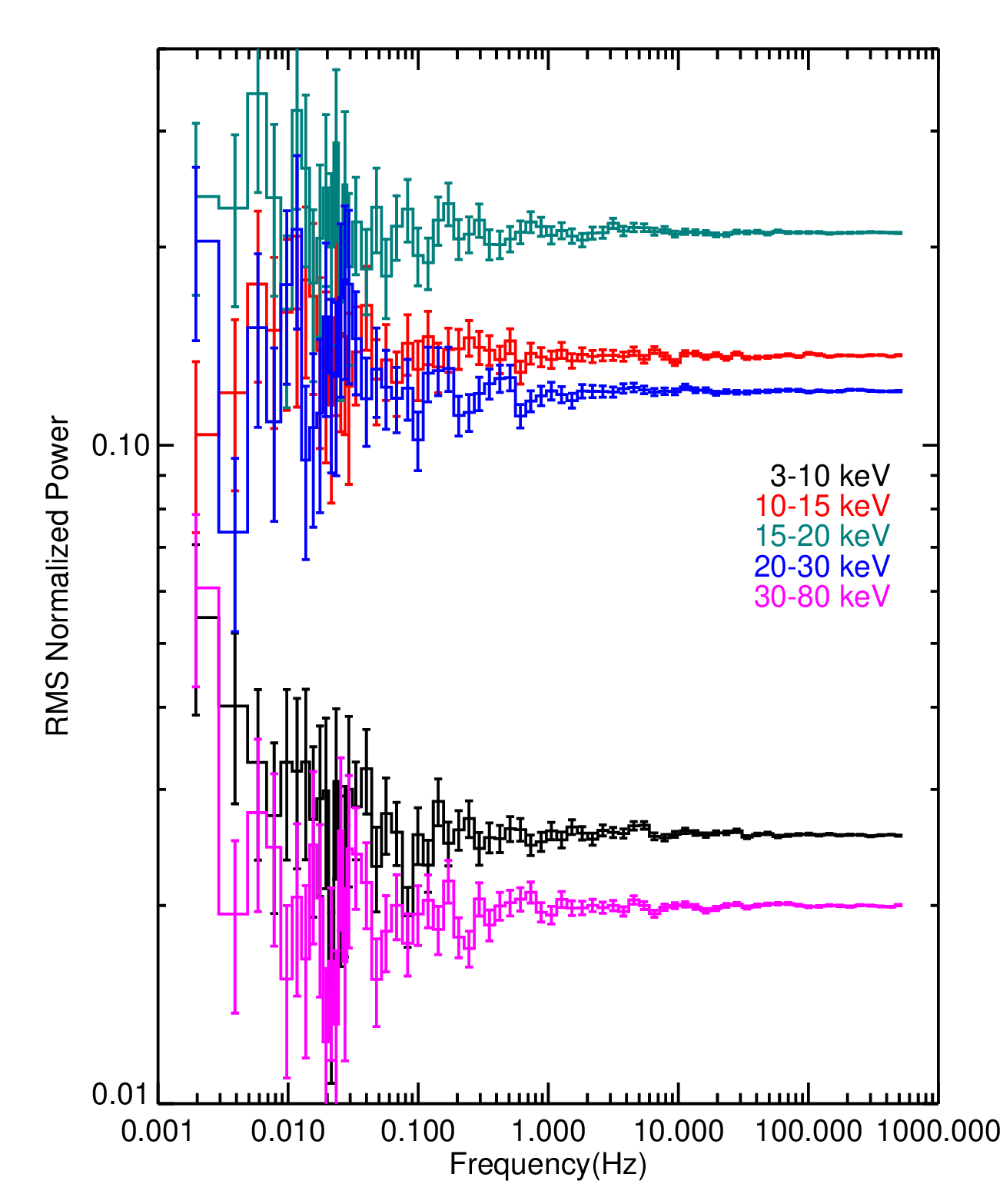}
    \caption{Logarithmically-rebinned power spectra (without Poisson noise subtractions) for observation 1 (February 27) in the RMS normalized power representation of 4U 1724--30 with a binfactor 1.2 (see Section \ref{section3.3}).  } 
     \label{eng_dep_lxp}
\end{figure}
Furthermore, we examine the energy-dependent variability in the 3-10 keV, 10-15 keV, 15-20 keV, 20-30 keV, and 30-80 keV energy bands for all four observational epochs. To study the nature of variability, we consider the same model as before, and all the power spectra are observed to be well described by the constant plus power-law $(\alpha+\beta \times (f/0.01)^{-\nu})$ model. Figure~\ref{eng_dep_lxp} shows the logarithmically-rebinned energy-resolved RMS normalized power spectrum for the observation 1 (LAXPC20). The RMS normalized powers are obtained by scaling the time-averaged Leahy power spectra by the intensity. Such a normalization gives a direct estimate of the variability present in the data. The nature of the energy-resolved power spectra is observed to be consistent for all four observations. The total integrated Root Mean Square (RMS) variabilities of the power spectra obtained for all four LAXPC observations are 6.22\%, 5.70\%, 5.38\%, and 7.64\%, respectively. To study the energy dependence of RMS variabilities, we obtain the fractional RMS considering different energy bands, namely 3-10 keV, 10-20 keV, 20-30 keV, and 30-80 keV, and we report them in Table~\ref{rms_var}. To obtain the fractional RMS, we follow the method mentioned in \citet{1990A&A...230..103B}.

\begin{figure}
    \centering
    \hspace*{-0.6cm}
    \includegraphics[width=0.50\textwidth]{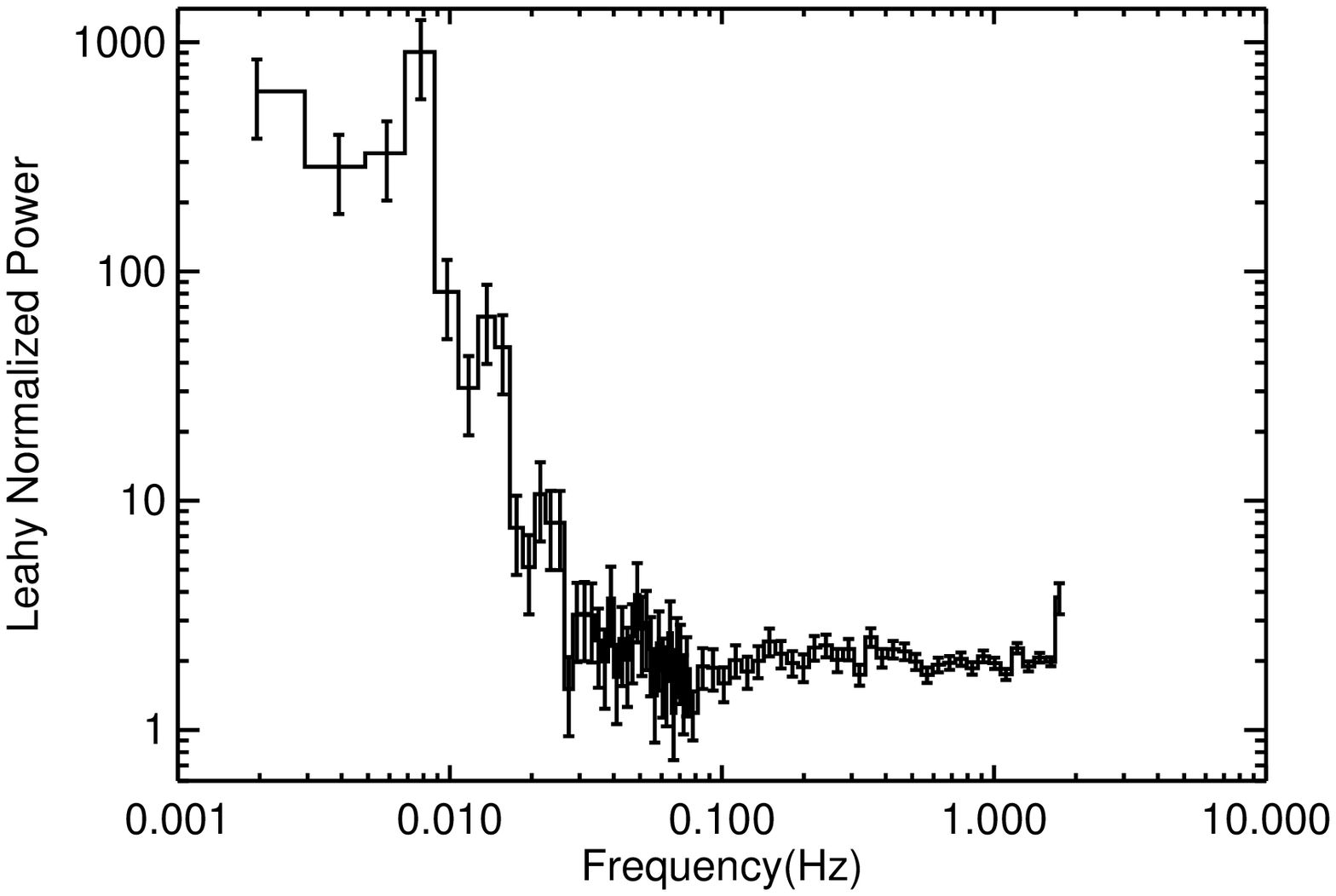}
    \caption{Logarithmically-rebinned (SXT) power spectrum for observation 1 (February 27) of 4U 1724--30 in the Leahy normalized power representation (see Section \ref{section3.3}). } 
    \label{pwspec_sxt}
\end{figure}

\begin{table}
\centering
\caption{RMS variability obtained using the method mentioned in section  \ref{section3.3} for the observation 1 (February 27), observation 2 (March 14), observation 3 (July 15) and observation 4 (July 29) of 4U 1724--30 for different energy bands (see Section \ref{section3.3}). }
\label{rms_var}
\begin{tabular}{|c|c|c|}
\hline
Obs No. &Energy (keV) & RMS Variability (\%)\\ 
\hline
1 & 3-10 & 0.42\\
&10-20 &0.60\\
&20-30&0.55\\
& 30-80&0.76\\
\hline

2 &3-10 &0.99\\
&10-20&1.33\\
&20-30 &1.66 \\
&30-80 &0.82\\
\hline
3 & 3-10 &0.84 \\
&10-20 &0.71\\
&20-30&1.66\\
&30-80&1.08\\
\hline

4 & 3-10 &0.95 \\
&10-20 &1.24\\
&20-30&1.44\\
&30-80 &1.39\\
\hline
\end{tabular}
\end{table}

Furthermore, we investigate the presence of frequency and energy-dependent time lags in the LAXPC data between different pairs of time series at different energies, considering the 3-10 keV as a reference energy band. We first compute the cross-spectrum considering integration time of 128 s and Nyquist frequency of 512 Hz, to calculate the phase lags for each energy band, from which time lags are obtained by dividing it by a factor $2 \pi f$, $f$ being the Fourier frequency \citep{1998astro.ph..7278N}. Our results show no significant variation in the time lag with energy or frequency for any observation.

Similar to LAXPC, with SXT data, we obtain the logarithmically rebinned Leahy normalized averaged power spectra using a Nyquist frequency of 1.7762 Hz and an integration time of 512 s for each of the observations \citep{1989ASIC..262...27V}. Figure~\ref{pwspec_sxt} shows the power spectrum obtained for observation 1 (SXT). We observe some unphysical periodic patterns in the lower frequency range of the SXT power spectrum, which is possibly due to satellite jitter also reported in \citet{2019MNRAS.488..720B}.  We try to fit the SXT power spectra with both constant plus power-law $(\alpha+\beta \times (f/0.01)^{-\nu})$ and constant $(\alpha)$ models, but none of these models are observed to provide an acceptable fit. It is possibly due to the atypical patterns observed in the lower frequency range of the power spectrum (Figure~\ref{pwspec_sxt}), and we are limited by the time-resolution of SXT ($\sim$ 278 ms) for further investigations of the shape of the power spectra towards the very low frequencies.

\section{Discussion}
\label{section4}

In this work, we present a broadband semi-simultaneous study of the poorly studied dim persistent NS LMXB 4U 1724--30 using LAXPC and SXT on board {\em AstroSat} over four observational epochs. We report the broadband spectral (0.7-25 keV) and timing (0.7-80 keV) analysis of the source. The hardness-intensity diagram of the LMXB 4U 1724--30 (Figure~\ref{hid}) shows that it was in the hard island state of its atoll track during all the observational epochs \citep{2008ApJ...687..488A} while being in a relatively softer state during observation 2.

\subsection{Spectral Properties}
\label{section4.1}
We investigate the joint broadband spectra of NS LMXB 4U 1724--30 in the 0.7-25 keV energy band using LAXPC and SXT on board {\em AstroSat}. The joint spectra obtained for all the observations are well modelled by the absorbed blackbody radiation along with a non-thermal emission (Comptonization) model. Typically, the origin of the soft thermal blackbody component is proposed to be centrally located and connected to the boundary layer (BL) close to the NS surface or the accretion disc or NS surface itself \citep{1998A&A...339..802G}. The detection of a Comptonization component with low temperature seed photons in our analysis suggests that the source of these photons is a cold, most likely truncated \citep{1994ApJ...434..570T, 2011MNRAS.411.2717M} accretion disc. Furthermore, the non-requirement of an additional disc blackbody component in the best-fit model of our analysis corroborates such interpretation. The derived blackbody temperatures (kT) from the fits are observed to vary from 1.41 keV to 1.66 keV, and the inferred size of the emitting region (radius $\sim$ 1.14-1.55 km) from the blackbody normalizations (1.85-3.40) obtained from our analysis for a source distance of $\sim$ 8.42 kpc (estimated using the peak PRE burst flux, see Section \ref{section4.2}) suggests that it is possibly a hotspot present on the NS surface itself. Our results suggest that the low-energy (soft) seed photons (temperature T0 $\sim$ 0.22-0.26 keV) possibly coming from a truncated, cold disc undergoes Comptonization by the surrounding hot corona (electron temperature $kT_{e}$ $\sim$ 8 keV and optical depth $\tau$ $\sim$ 2.62-3.77). From the evolution of the parameters (Figure~\ref{para_evol}), it is observed that the galactic neutral hydrogen column density $n_H$, blackbody temperature $kT$, Comptonization normalization, and source flux are relatively higher during the second observation as compared to the other three observations. The variation of the spectral parameters corresponding to observation 2 is consistent with the relatively softer nature of the source in this observation, which is observed in the HID as well. Physically, for observation 2, the higher source flux suggests an increasing accretion rate (Figure~\ref{para_evol}). In the relatively softer state, the disc may approach closer to the central compact object or the boundary layer, thereby further heating the accretion flow, possibly resulting in the relatively higher thermal emission ($kT \sim 1.66$ keV) from the disc-boundary layer. The higher value of Comptonization normalization may indicate increasing Comptonization efficiency, leading to the variation in the optical depth of the hot plasma during observation 2. In \cite{1998A&A...339..802G}, the high-energy spectrum of 4U 1724--30 was reported to be well interpreted by a {\tt comptt} model (non-thermal component) describing Comptonization of soft seed photons ($kT \sim 1$ keV) by a spherically-symmetric hot plasma ($kT_e \sim 30$ keV) of optical depth ($\tau$ $\sim$ 3). While a boundary layer between the accretion disc and the compact object was reported to be one of the possible origins of the thermal component. However, the deviation in the inferred size of the thermal emitting region with our result is possibly due to the relatively lower luminosity ($\sim$ $10^{36}$ erg/s) harder source state during the {\em AstroSat} observations.

\subsection{The Burst}
\label{section4.2}
A Type-I (thermonuclear) X-ray burst with double-peaked morphology \citep{1998A&A...339..802G, 1980ApJ...240L.121G, 2011ApJ...742..122S} is detected in the LAXPC light curve of observation 4. The burst spectra extracted over 2 s segments are well described by the variable persistent emission model and the maximum temperature, flux, and radius are 2.87 keV, $4.25 \times 10^{-8}$ $\mathrm{erg\,s^{-1} cm^{-2}}$ and 14.46 km respectively (Figure~\ref{burst_para_evol}). The variation of the blackbody radius ($R$) inferred from the blackbody normalization ($N$) as well as blackbody temperature ($kT$) variations near the burst peak suggest the PRE nature of the burst \citep{2018SSRv..214...15D}. The detections of PRE bursts have also been reported previously from 4U 1724--30 \citep{2008ApJS..179..360G,2000A&A...357L..41M,2011ApJ...742..122S}. Considering the Eddington flux of 4U 1724--30 as $4.25^{+0.23}_{-0.22} \times 10^{-8}$ $\mathrm{erg\,s^{-1} cm^{-2}}$ ($\sim$ 1.78 crab) as obtained from our burst analysis, the distance is estimated as $\sim$ $8.42^{+0.33}_{-0.32}$ kpc, assuming Eddington luminosity of  $\sim$ 3.79 $\times 10^{38}$ $\mathrm{erg\,s^{-1}}$ \citep{2003A&A...399..663K}. The touch-down flux and distance estimations based on the peak PRE burst flux are consistent with the previously reported estimates of the X-ray burster 4U 1724-30 \citep{2020ApJS..249...32G}. Furthermore, the distance estimate based on our result is consistent with the previously reported distance limit of the source \citep{2003A&A...399..663K,1996yCat.7195....0H}. However, the estimated source distance is not exactly consistent with the distance ($\sim$ 7.2 kpc) reported in \cite{2000AIPC..510..217C}, and the reasons behind such deviation are possibly the uncertainties of peak burst flux estimate ($\sim$ 15\%), and uncertainties in Eddington luminosity \citep{2003A&A...399..663K,Galloway_2003,2008ApJS..179..360G}. These uncertainties may explain the discrepancy between the distance value obtained in this work and the measure by \cite{2000AIPC..510..217C} (see \cite{marino:hal-02371621}, for further discussion).

Additionally, the properties of the burst detected in our data such as observed burst peak flux ($\sim 4.25 \times 10^{-8}$ $\mathrm{erg\,s^{-1} cm^{-2}}$), burst duration ($\sim$ 50 s) as well as low accretion rate inferred from the continuum spectra (persistent flux $\sim 10^{-10} $ $\mathrm{erg\,s^{-1} cm^{-2}}$ ) implies that it is possibly a hydrogen burst \citep{2008ApJS..179..360G}. 
  
The evolution of $f_{a}$ (bottom right panel of Figure~\ref{burst_para_evol}) can be interpreted as the increase in the mass accretion rate onto the NS during the burst \citep{2018ApJ...860...88B}. This increase may indicate the effects of burst radiation-induced Poynting-Robertson (PR) drag on the disc material  \citep{2013ApJ...772...94W, 2015ApJ...806...89J,2010A&A...520A..81I,2020NatAs...4..541F,10.1093/mnras/stab3087}.
Such burst feedback on the persistent emission from the accretion disc has also been observed for other sources \citep[e.g., Rossi X-ray Timing Explorer (RXTE) observation of GS 1826-238; ][]{2015ApJ...806...89J}, and is an example of burst-accretion interaction.

Furthermore, from the energy-dependent burst profile (Figure~\ref{burstlc}), it is observed that the average hard X-ray (30-80 keV) profile varies significantly from the energy integrated (3-80 keV) light curve during the burst. Near the burst peak, a significant shortage in hard X-rays (30-80 keV counts) is observed, and this suggests coronal cooling during the burst peak. During the thermonuclear ignition, the hot plasma undergoes Comptonization by the incident softer burst seed photons from the NS surface. Hence, the energy of the corona is expected to decrease, resulting in a decrease of the coronal temperature \citep{2014ApJ...782...40J}. The soft photons from the thermonuclear burst may accelerate the coronal cooling process, leading to the hard photon shortage near the peak of the burst \citep{10.1093/mnras/staa3137,2012ApJ...752L..34C, 2003A&A...399.1151M}. Such shortage in hard X-ray photon counts is generally observed in the hard state of LMXBs \citep{2014ApJ...782...40J}. This is consistent with the spectra, which reveal the source 4U 1724--30 to be in the low-luminosity non-thermally dominated (hard/island)  state over the four observational epochs (Figure~\ref{spec_obs3}). As the burst decays, the hard photon counts recover original values gradually (Figure~\ref{burstlc}). Thus, the burst emission has a significant influence on the corona. Similar kinds of shortage in the hard X-ray count rates have been reported previously during burst detected by RXTE from the clocked burster GS 1826--238  \citep{2015ApJ...806...89J, 2014ApJ...782...40J,2020A&A...634A..58S}.

\subsection{Timing Studies}
\label{section4.3}
Variability properties and Quasi-Periodic Oscillations (QPOs) at different frequencies have previously been reported from the NS atoll source 4U 1724--30  \citep{2008ApJ...687..488A, 1998A&A...333..942O, 1999NuPhS..69..245B}. In our analysis, we have done a detailed investigation of the shape of the power spectra (Figure~\ref{pwspec_lxp} and Figure~\ref{pwspec_sxt}) for the presence of any features in addition to the Poisson noise, and we do not detect any narrow features (QPOs). The 1 $\sigma$ upper limits on any possible periodicity in the frequency range 100-500 Hz are mentioned in Table~\ref{modl_para}. The averaged normalized  power spectra (LAXPC) are observed to comprise of low-frequency (below $\sim 0.1$ Hz) noise (red noise) described by a power-law (Table~\ref{modl_para}) in addition to the usual Poisson noise.
The presence of low-frequency noise indicates the presence of fast variabilities. These fast variabilities may be affected by the accretion process as well as the low magnetic field of the system \citep{2003A&A...398.1103R}. The presence of low-frequency power-law noise in the power spectra reveals inhomogeneous accretion flow, which may occur due to various mechanisms that may result in clumping in the disc-boundary layer in weakly magnetized NS.

The shapes of the LAXPC power spectra at different energies are observed to exhibit a similar pattern with the presence of noise (red noise) at low frequencies following a power-law (Figure~\ref{eng_dep_lxp}). The variability amplitudes initially show an increasing trend with energy and drop at higher energy bands for all four observational epochs (Figure~\ref{eng_dep_lxp}). Table~\ref{rms_var} also shows that the RMS variabilities obtained for the corresponding 4 energy bands (LAXPC) exhibits a positive correlation with energy for the first three energy bands (3-10 keV, 10-15 keV, 15-30 keV) and then observed to decrease again at higher energies. Such correlations were observed from this source previously as well \citep{1998A&A...333..942O}. Typically, at higher energy, the non-thermal emission dominates. The lower values of variability amplitudes along with the low fractional RMS (Table~\ref{rms_var}) in the higher energy bands may imply that the soft-seed photon variability is dominant over the Comptonization fluctuations. The propagation of fluctuations in mass accretion rate may also give rise to the observed variability \citep{2013MNRAS.434.1476I}. Alternatively, if the coronal fluctuation is giving rise to the observed variability, in that case, the decrease in variability at the high energy can be due to the non-thermal acceleration in the hot plasma \citep{1990ApJ...357..149Z,Bu_2021}. This effect is dominated by Compton cooling, resulting in decreased variability in the higher energy range \citep{10.1111/j.1365-2966.2005.09527.x}. It should be noted here that that there may also be contributions from the higher background and lower effective area of LAXPC at such high energies. Furthermore, the presence of a low-frequency noise component in the power density spectra is a characteristic of NS atoll sources and is often described by a power-law (``$1/f^{\nu}$ noise"). Additionally, the low-frequency noise of these sources is observed to be weaker with typical values $\leq$ 1 (RMS) in the hard island state \citep{1989A&A...225...79H}. Thus, the results presented in this work manifest that the source 4U 1724-30 shows timing behavior similar to other NS atoll LMXBs.

In general, time lags between variabilities at different energies can be interpreted physically as the transfer of fluctuations within the accretion disc. It can also be the fluctuation propagation between the disc and the corona, the source of the high energy component where the soft seed photons from the disc get compromised \citep{2011MNRAS.414L..60U}. The energy-dependent lags or frequency-dependent lags obtained from our analysis are not significant and observed to be consistent with a zero-lag considering $1 \sigma$ uncertainty. The reason behind the absence of such observed lag may possibly be due to the same emitting (spectral) component dominating across the different energy bands considered. This implies the significant hard non-thermal emission ({\tt comptt}) dominating state of the NS LMXB, which is also evident from the joint spectra. The absence of lags, therefore may reflect the significant overlap between the non-thermal emission-dominated regions across different energy bands \citep{2012MNRAS.427.2985C}.

\section{Conclusions}

 Despite intensive studies of the LMXBs in X-rays, broadband investigations of dim LMXBs with good sensitivity, which is vital for unveiling the binary emission components and their evolution at low accretion rates, are so far limited.
Our LAXPC and SXT on board {\em AstroSat} observation have enabled a detailed broadband study of the dim persistent NS LMXB 4U 1724--30. The source was observed in the hard island state of its atoll track, exhibiting a modest spectral evolution. The X-ray energy spectra show the hard non-thermal emission dominated state of the source during all four observations. The timing analysis reveals the presence of low-frequency noise in the averaged power spectra. This kind of study of dim LMXB sources provides scopes to understand the physics of accretion at a low accretion rate. The time-resolved spectroscopy of the LAXPC detected Type-I burst reveals the expansion and contraction of the NS photosphere near the Eddington limit of the source, thereby probing its PRE nature. The hard X-ray shortage and the enhanced accretion flow observed during the burst reveal the presence of burst feedback on the overall accretion processes. Such a study helps to get an insight into the nature of physical processes occurring in the accretion flow and the corona, and the correlations of the burst property on the spectral states of such binary systems. A similar kind of spectro-timing analysis in the X-ray regime using {\em AstroSat}, as well as simultaneous observation using other new-generation telescopes with better soft X-ray sensitivity such as {\em NICER}, will provide a better scope for more detailed studies, including flux evolution, spectral evolution, time-resolved spectroscopy as well as energy-dependent variability of such low luminosity systems.

\section*{Data Availability}

This paper includes data publicly available from the Indian Space Science Data Centre (ISSDC), ISRO website [\url{https://astrobrowse.issdc.gov.in/astro\_archive/archive/Home.jsp}].

\section*{Acknowledgements}

We are grateful for the referee's constructive comments and suggestions. This publication uses the data from the {\em AstroSat} mission of the Indian Space Research Organisation (ISRO), archived at the Indian Space Science Data Centre (ISSDC). This work particularly uses data from the Soft X-ray Telescope (SXT), and the Large Area X-ray Proportional Counter (LAXPC) developed mainly at TIFR, Mumbai, and the SXT and LAXPC Payload Operation Centres at TIFR are thanked for verifying and releasing the data via the ISSDC data archive and providing the necessary software tools. We thank Tomaso M. Belloni for providing the \texttt{GHATs} timing package, which helped us verify some of our results. We acknowledge support from ISRO under the AstroSat archival
Data utilization program (DS 2B-13013(2)/4/2020-Sec.2).




\bibliographystyle{mnras}
\bibliography{ref_1724.bib} 





\appendix




\bsp	
\label{lastpage}
\end{document}